\documentclass[pre,twocolumn,amsmath,amssymb,showpacs]{revtex4-1}
\usepackage{graphicx,psfrag,color,hyperref}
\usepackage{amssymb,latexsym,mathrsfs,amsmath,mathrsfs}
\begin{document}
\def\be{\begin{equation}}
\def\ee{\end{equation}}
\def\bea{\begin{eqnarray}}
\def\eea{\end{eqnarray}}

\title{Diffusion with stochastic resetting at power-law times}
\author{Apoorva Nagar$^1$ and Shamik Gupta$^2$}
\affiliation{\mbox{$^1$Indian Institute of Space Science and Technology, Thiruvananthapuram, Kerala, India}\\ \mbox{$^2$Max Planck Institute for the Physics of Complex Systems, Noethnitzer Stra{\ss}e 38, D-01187 Dresden, Germany}}
\begin{abstract}
What happens when a continuously evolving stochastic process is interrupted with large changes at random intervals $\tau$ distributed as a power-law $\sim \tau^{-(1+\alpha)};\alpha>0$? Modeling the stochastic process by diffusion and the large changes as abrupt resets to
the initial condition, we obtain {\em exact} closed-form expressions for both static and dynamic quantities, while accounting for strong correlations implied by a power-law. Our results show that the resulting dynamics exhibits a spectrum of rich long-time
behavior, from an ever-spreading spatial distribution for $\alpha < 1$, to one that is
time independent for $\alpha > 1$. The dynamics has
strong consequences on the time to reach a distant target for the first time; we specifically show that there
exists an optimal $\alpha$ that minimizes the mean time to
reach the target, thereby offering a step towards a viable strategy to
locate targets in a crowded environment. 
\end{abstract}
\date{\today}
\maketitle

A wide variety of physical phenomena during evolution undergo sudden {\em large}
changes over a time substantially shorter
than the typical dynamical timescale, e.g., financial crashes due to
fall in stock prices~\cite{Sornette:2003}, sudden reduction in
population size due to catastrophes \cite{Newman:2003}, and sudden
changes in tectonic plate location in earthquakes.
Often the time series of these phenomena exhibits bursts of
intense activities separated by intervals distributed as a
power-law, e.g., in earthquakes \cite{Bak:2002}, material failure under
load fatigue \cite{Kun:2008}, coronal mass ejection from the
sun \cite{Lippiello:2008}, fluorescence decay of nanocrystals and
biomolecules \cite{Chicos:2007,Kierdaszuk:2010}, neuron
firings \cite{Kemuriyama:2010}, successive
crashes in stock exchanges \cite{Gontis:2007,Sornette:2003,
Lillo:2003}, and email sending times \cite{Barabasi:2005}. Considering the underlying generic situation of a continuously evolving process interrupted by sudden large changes at random times, a pertinent question of theoretical and practical relevance is then: How do these interruptions affect the observable properties at long times? To get a first answer, one may model the continuously evolving process by the widely relevant example of diffusion, and the large changes as resets to the initial state. 

Diffusion with stochastic resetting has been extensively studied in recent times. Starting with a single diffusing particle resetting to its initial position \cite{Evans:2011-1,Majumdar:2015-1}, subsequent works studied motion in a bounded domain \cite{Christou:2015}, in a potential \cite{Pal:2015}, for many choices of resetting position \cite{Evans:2011-2,Boyer:2014,Majumdar:2015-2}, for a continuous-time random walk \cite{Montero:2013,Mendez:2016}, for L\'{e}vy \cite{Kusmierz:2014} and exponential constant-speed flights \cite{Campos:2015}. Resetting was also studied in interacting particle systems such as fluctuating interfaces \cite{Gupta:2014,Majumdar:2015-1} and reaction-diffusion models \cite{Durang:2014}. Diffusion combined with stochastic resetting mimics the natural search strategy, whereby an unsuccessful search continues by returning to the starting position \cite{Evans:2011-1}, and was used to optimize search in combinatorial problems \cite{Lovasz:1996,Montanari:2002,Konstas:2009}. A naturally occurring example of resetting in many-particle systems is during protein production by ribosomes moving on mRNA, when the latter suddenly degrades at random times and the dynamics resets to the initial condition  with the production of a new mRNA \cite{Nagar:2011,Valleriani:2010,Valleriani:2011}.

While the above works considered resetting at exponentially-distributed
times (or, a generalized exponential \cite{Eule:2016}), we consider here
a power-law distribution. Even with random walks, changing the waiting time
distribution for jumps from an exponential to a power-law leads to significant consequences, e.g., rendering normal diffusion anomalous \cite{Metzler:2000,Metzler:2004,Klages:2008}; we may then already
anticipate our model with a power-law instead of an exponential for
resetting times to result in dramatic changes. Diffusion involves spreading out of a dynamical observable from a region of high to low
concentration, which in the absence of boundaries continues for all times. In presence of resetting, the
opposing tendencies of diffusive spreading and confinement around the
initial state due to the abrupt resets lead to surprisingly rich behaviors. As the exponent of the power-law varies, the change in the
relative dominance of diffusion {\em vis-\`{a}-vis} resetting results in
significantly different behaviors. Strong correlations implied by a power law pose a challenge for analytic
tractability, yet, remarkably, we are able to characterize these
multiple behaviors by {\em exact} closed-form expressions for both static and dynamic quantities.

\begin{figure*}
\centering
\includegraphics[width=15cm]{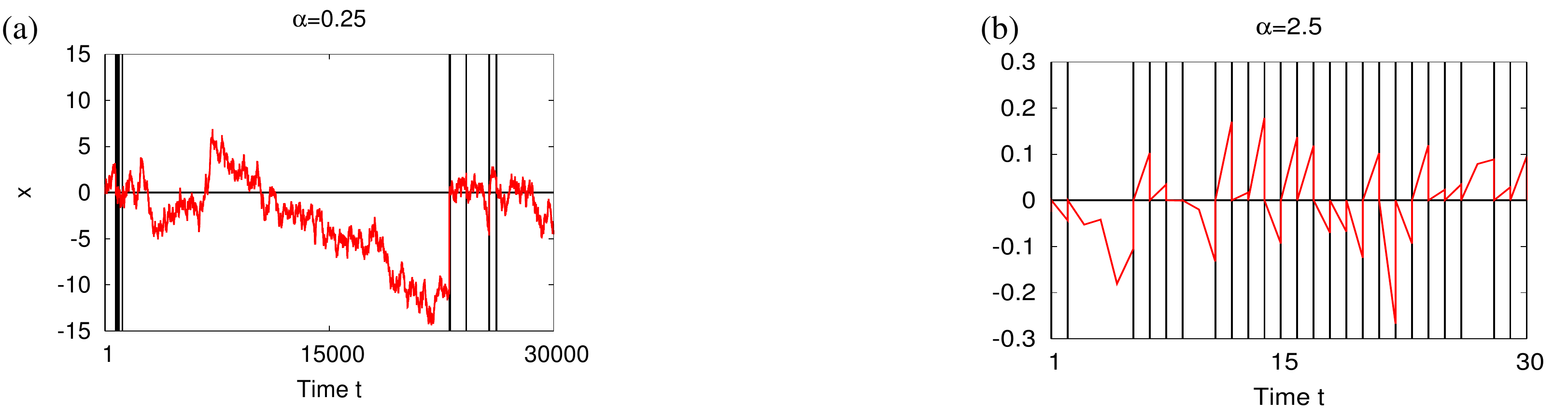}
\caption{Typical space-time trajectories (red lines),
with black lines marking resetting events: Resetting location $x_0=0$, diffusion constant $D=0.5$, $\tau_0=1.0$.
}
\label{fig:full-picture}
\end{figure*}

\begin{figure*}
\centering
\includegraphics[width=15cm]{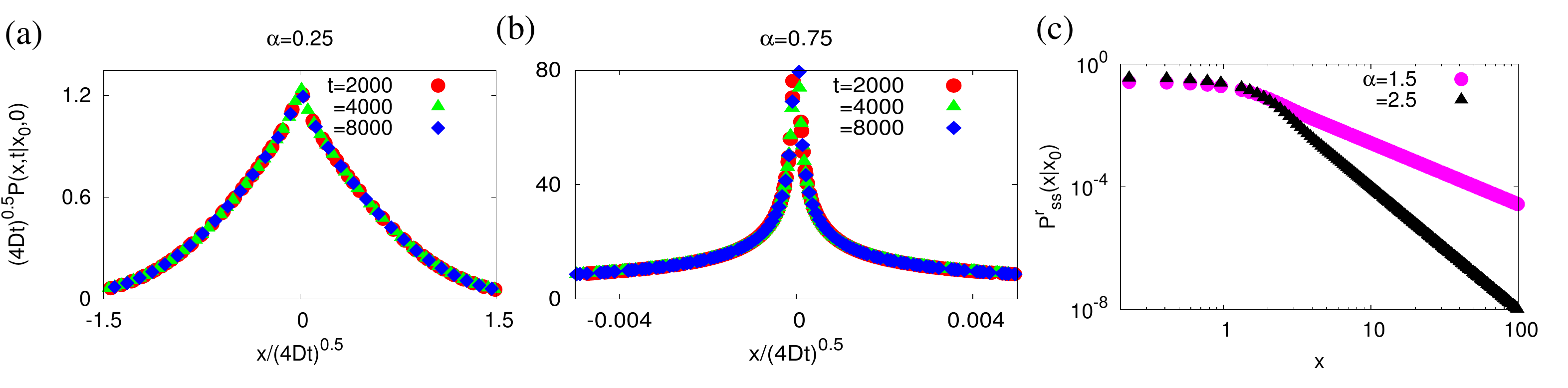}
\caption{(a),(b): Data collapse of exact spatial distribution for
$\alpha<1$ for different times, following Eq.
(\ref{eq:pxt-alphalt1-exact-1}). (c) Time-independent distribution for
$\alpha>1$, Eq. (\ref{eq:ss-distr}). Resetting location $x_0=0$, diffusion constant $D=0.5$, $\tau_0=1.0$.}
\label{fig2}
\end{figure*}

In this work, we consider a particle with diffusion constant $D$ diffusing
in one-dimension $x$, and being interrupted at random times by a reset to its initial location $x_0$. The time $\tau$ between successive resets is distributed as
a power-law:
\be
\rho(\tau)=\frac{\alpha}{\tau_0(\tau/\tau_0)^{1+\alpha}};~\tau\in[\tau_0,\infty),~\alpha>0,
\label{eq:ptau}
\ee
with $\tau_0$ a microscopic cut-off. Figures \ref{fig:full-picture}(a),(b) show typical space-time trajectories for
representative $\alpha$'s. Note that for $\alpha<1$, all moments of
$\rho(\tau)$ are infinite. For $\alpha>1$, the first moment is finite: $\langle \tau \rangle=\tau_0 \alpha/(\alpha-1)$, while
for $\alpha>2$, the second moment also becomes finite: $\langle
\tau^2 \rangle=\tau_0^2\alpha/(\alpha-2)$. By contrast, the previously-studied exponential $\rho(\tau)$ always has finite mean and variance.  
Also, an exponential $\rho(\tau)$ implies a resetting at any time
to occur with a constant probability. By contrast, a power-law
distribution implies, depending on $\alpha$, the corresponding
probability to depend explicitly on time.

Our exact results for the long-time properties of the system show
that the spatial probability distribution exhibits on tuning
$\alpha$ a rich behavior with multiple crossovers. For $0<\alpha<1$,
the average gap $\langle \tau \rangle$ between successive resets being
infinite, a typical space-time trajectory in a given time has a small
number of reset events, and in between diffuses further away from the initial location, Fig.
\ref{fig:full-picture}(a); this leads to a
spatial distribution with a width that continually increases in time as
$\sqrt{t}$, similar to diffusive spreading. The behavior for $\alpha<1$ is captured in the
scaling plots in Figs. \ref{fig2}(a),(b). By contrast, for $\alpha>1$, a finite $\langle \tau \rangle$
implies frequent resets in a given time, so that the particle does not diffuse too far from its initial location, Fig.
\ref{fig:full-picture}(b). Hence, one has at long times a spatial
probability distribution that no longer spreads in time, but is time
independent with power-law tails (Fig. \ref{fig2}(c)); nevertheless,
fluctuations as characterized by the mean-squared
displacement (MSD) diverge with time for $1<\alpha<2$, while a time-independent behavior emerges only for $\alpha>2$. Previous
studies for an exponential $\rho(\tau)$ have shown that
diffusion with resetting always leads to a
time-independent spatial distribution with a finite MSD. Our work highlights that such a scenario does not necessarily hold for a power-law $\rho(\tau)$.

Besides the crossovers at $\alpha=1,2$, there is another one at $\alpha=1/2$, where the time-dependent spatial distribution near the resetting location changes over from a cusp for $0<\alpha<1/2$ (Fig. \ref{fig2}(a)) to a
divergence for $1/2<\alpha<1$ (Fig. \ref{fig2}(b)). This feature may be
contrasted with exponential resetting, where the spatial
distribution at long times always exhibits a cusp singularity \cite{Evans:2011-1}.
As we will show, this difference in behavior is linked to
resetting events occurring with a probability that is time independent
for an exponential $\rho(\tau)$, but
which has an essential time dependence for a power-law $\rho(\tau)$ for
$0<\alpha<1$.
We also study the mean first passage time (MFPT) for the
diffusing-resetting particle to reach a distant target fixed in space. The MFPT is an important
quantifier of practical relevance, e.g., for a diffusing reactant on a
polymer that has to react with an external reactive site fixed in space
\cite{Guerin:2012,Guerin:2013}. A surprise emerging from our results is that for $\alpha>1$, the MFPT
exhibits a non-monotonic dependence on $\alpha$, implying an
optimal $\alpha$ that minimizes the MFPT to reach a given
target. The derivation and
understanding of these results constitute the rest of this paper.  

We begin with deriving $P^{{\rm r}}(x,t|x_0,0)$, the
probability density for the particle to be at $x$ at time $t$, 
given $x=x_0$ at $t=0$. This probability depends solely on trajectories
originating at the last reset prior to $t$, when the motion starts
afresh (gets ``renewed") at $x_0$. Then, $P^{{\rm r}}(x,t|x_0,0)$
is given by the propagator
$P(x,t|x_0,t-\tau)\equiv\exp[-(x-x_0)^{2}/(4D\tau)]/\sqrt{4\pi D\tau}$ of free
diffusion for time $\tau$ ($\tau\in[0,t]$) elapsed since the last reset, weighted by
the probability density $f_\alpha(t,t-\tau)$ at time $t$ for the last
reset to occur at time $t-\tau$, as \cite{norm-note}
\be
P^{{\rm r}}(x,t|x_0,0)=\int_0^{t}d\tau f_\alpha(t,t-\tau)P(x,t|x_0,t-\tau).
\label{eq:pxt-gen}
\ee

To proceed, we require $f_\alpha(t,t-\tau)$, which is given by the probability density $G(t-\tau)$ for a reset at time $t-\tau$ and the probability $\rho_0(\tau)$ for no reset in the interval $[t-\tau,t]$, as
$f_\alpha(t,t-\tau)=\rho_0(\tau)G(t-\tau)$, where $\rho_0(\tau)\equiv \int_\tau^\infty d\tau'\rho(\tau')=(\tau/\tau_0)^{-\alpha};~\tau \ge
\tau_0$, using Eq. (\ref{eq:ptau}). Let $g_n(t);n\ge0$, be the probability density for the $n$-th reset at
time $t$, with $\int_0^\infty dt g_n(t)=1~\forall~n$. Here,
$g_0(t)=\delta(t)$ accounts for the initial condition $x=x_0$ at $t=0$,
which itself is a reset. One has
\cite{Cox:1962} $g_n(t)=\int_0^{t}d\tau\rho(t-\tau)g_{n-1}(\tau);~n\ge1$,
since the probability for the $n$-th reset at time $t$ is given by the
probability for the $(n-1)$-th reset at an earlier time $\tau$ and the probability that
the next reset happens after an interval $t-\tau$. By definition, we have
$G(t)=\delta(t)+\sum_{n=1}^{\infty}g_n(t)$, and a straightforward
calculation using Laplace transform (LT) to compute
$g_n(t)$ yields for large $t$ that $G(t)=1/\langle \tau \rangle$
for $\alpha>1$, and $G(t)=t^{\alpha-1}$ for $0<\alpha<1$. For an exponential $\rho(\tau)=r\exp(-r\tau)$, $G(t)=r$ for all $t>0$. By contrast, for the power-law for $0<\alpha<1$, $G(t)$ is time dependent,
which we show later to significantly affect the observable properties.
We get for $t \gg \tau_0$ \cite{SM,Godreche:2001}
\bea
&&f_{\alpha<1}(t,t-\tau)=\frac{\sin(\pi\alpha)}{\pi}\tau^{-\alpha}(t-\tau)^{\alpha-1},
\label{eq:fB-alphalt1} \\
&&f_{\alpha>1,\tau \ge \tau_0}(t,t-\tau)=
\frac{1}{\tau_0}\Big(\frac{\alpha-1}{\alpha}\Big)\Big(\frac{\tau}{\tau_0}\Big)^{-\alpha},
\label{eq:fB-agt1}
\eea
and $\int_0^{\tau_0}d\tau
f_{\alpha>1,\tau<\tau_0}(t,t-\tau)=1-\int_{\tau_0}^t
d\tau~f_{\alpha>1,\tau \ge \tau_0}(t,t-\tau)$. Knowing $f_\alpha$, Eq. (\ref{eq:pxt-gen}) allows to derive $P^{\rm r}(x,t|x_0,0)$.

{\bf Spatial distribution, $\alpha<1$}: For large $t \gg \tau_0$, we have \cite{SM}
\be
P^{\rm r}(x,t|x_0,0)=\frac{\Gamma(\alpha)\sin(\pi
\alpha)e^{-\frac{z}{t}}}{\pi\sqrt{4\pi
Dt}}U\left(\alpha,\alpha+\frac{1}{2},\frac{z}{t}\right),
\label{eq:pxt-alphalt1-exact-1}
\ee
where $z\equiv (x-x_0)^2/(4D)$, and $U(a,b,x)$ is the confluent Hypergeometric
function \cite{Abramowitz:1972}.
In the limit $t \to \infty$, the right hand side does not approach
a time-independent form. Since the average time $\langle \tau \rangle$
between successive resets is infinite for $\alpha<1$, a typical
space-time trajectory shows bursts of resets separated by
very long time intervals during which the particle
diffuses further and further away from its initial position, see Fig.
\ref{fig:full-picture}(a), leading to the spatial distribution
(\ref{eq:pxt-alphalt1-exact-1}) that continually broadens in time. While
$\langle x-x_0 \rangle=0$ due to the mirror symmetry about $x_0$ of the dynamics, the MSD grows linearly with time as in pure diffusion. The
time dependence in Eq. (\ref{eq:pxt-alphalt1-exact-1}) is captured by
the data collapse in Figs.
\ref{fig2}(a),(b). 

The limiting behavior of $P^{\rm r}(x,t|x_0,0)$ for small and large $x$
reveals rich and hitherto unexpected features. Using large and small $x$
behavior of $U(a,b,x)$ \cite{U-note} yields
\bea
P^{\rm r}(x,t|x_0,0) \sim 
\begin{cases}
&
\frac{\Gamma(\alpha-1/2)}{(4Dt)^{1-\alpha}}\frac{\sin(\pi\alpha)}{\pi^{3/2}|x-x_0|^{2\alpha-1}};
\\&|x-x_0|\to 0,\frac{1}{2}<\alpha<1,\\
& \frac{\Gamma(1/2-\alpha)\Gamma(\alpha)}{\sqrt{4\pi
Dt}}\frac{\sin(\pi\alpha)}{\pi^{3/2}};\\&|x-x_0|\to 0,\alpha<\frac{1}{2}, \\
& e^{-(x-x_0)^2/(4Dt)};|x-x_0| \to \infty.
\end{cases}
\label{eq:a<1limits}
\eea
Thus, as $|x-x_0|\to 0$, the behavior crosses over from being with a
cusp for $\alpha<1/2$ (Fig. \ref{fig2}(a)) to being divergent for $1/2<
\alpha<1$ (Fig. \ref{fig2}(b)).  This crossover behavior stems from the
form of $f_{\alpha<1}(t,t-\tau)$, which
is peaked at $\tau=0,t$, implying that most resets are close to either
the present or the initial time. However, as $\alpha$ crosses $1/2$, the
relative weight of these peaks changes, with the peak at $\tau=0$
becoming more dominant for $\alpha>1/2$; this
leads to a significant increase in reset events at small intervals prior
to the time of observation, thereby increasing the probability for the
particle to be close to the resetting location, and effecting the
mentioned crossover from a cusp to a divergence around $x_0$ across
$\alpha=1/2$.
The behavior of $P^{\rm r}(x,t|x_0,0)$ for $|x-x_0| \gg 1$ is dominated
by the propagator of the free diffusing particle, due to many
trajectories having last resets close to the initial time and free
diffusion without reset at subsequent times.

{\bf Spatial distribution, $\alpha>1$}: We get for $t \gg \tau_0$
\cite{SM}:
\bea
&&P^{\rm
r}(x,t|x_0,0)=\Big\{1-\frac{1}{\alpha}\Big[1-\Big(\frac{t}{\tau_0}\Big)^{1-\alpha}\Big]\Big\}\frac{\exp(-z/\tau_0)}{\sqrt{4\pi
D \tau_0}}\label{eq:pxt-agt1-full-eqn-1}\\
&&+
\frac{(\alpha-1)\tau_0^{\alpha-1}}{\alpha \sqrt{4\pi
D}}\Big[\frac{\gamma(\beta,z/\tau_0)}{z^{-\beta}}\nonumber 
 -\frac{e^{-\frac{z}{t}}}{t^{\beta}}\sum_{k=0}^{\infty}\frac{\Gamma(\alpha-1/2)(z/t)^k}{\Gamma(\alpha+k+1/2)}\Big],
 \nonumber
\eea
where $\beta\equiv\alpha-1/2$ and $\gamma(a,x)$ is the lower incomplete
Gamma function. As before, $\langle x-x_0 \rangle=0$ by symmetry, while the MSD for
$\alpha>2$ converges at long times to $2D\tau_0(\alpha-1)^2/(\alpha(\alpha-2))$,
 and diverges with time for $1<\alpha<2$ as $t^{2-\alpha}$, thus
 exhibiting a crossover at $\alpha=2$. 

Unlike for $\alpha<1$, here $P^{\rm r}(x,t|x_0,0)$ is independent of
time as $t\to\infty$ to yield a non-trivial steady state \cite{note}
\be
P^{\rm r}_{{\rm
ss}}(x|x_0)=\Big(\frac{\alpha-1}{\alpha\sqrt{4\pi
D\tau_0}}\Big)\mathcal{G}\Big(\frac{|x-x_0|}{\sqrt{4D\tau_0}}\Big);
\label{eq:ss-distr}
\ee
$\mathcal{G}(y)=y^{1-2\alpha}\gamma(\alpha-1/2,y^{2})+e^{-y^2}$.
Using $\gamma(a,x)/x^{a}\to1/a$ as $x\to0$, $\gamma(a,x)\to\Gamma(a)$
as $x\to\infty$ gives
\bea
P^{\rm r}_{{\rm
ss}}(x|x_0) \sim 
\begin{cases}
& \frac{(\alpha-1)(2\alpha+1)}{\alpha(2\alpha-1)\sqrt{4\pi
D\tau_0}};|x-x_0|\to 0,\\
& \frac{(\alpha-1)\Gamma(\alpha-\frac{1}{2})}{\alpha\sqrt{4\pi
D\tau_0}}\Big[\frac{4D\tau_0}{(x-x_0)^{2}}\Big]^{\alpha-1/2};|x-x_0| \to \infty.
\end{cases}
\label{eq:px-tails-alphagt1}
\eea
The steady state distribution has power-law tails and a cusp around $x_0$, Fig. \ref{fig2}(c).
Equation (\ref{eq:pxt-agt1-full-eqn-1}) implies a late-time relaxation to the
steady state as $\sim t^{1/2-\alpha}$. As for $\alpha<1$,
$f_{\alpha>1}(t,t-\tau)$ explains the above behavior: Eq. (\ref{eq:fB-agt1}) implies a large number of resets in the small interval $[t,t-\tau_0]$, while those outside this interval occur with a probability decaying as a
power-law. Hence, the probability of finding the particle very far from
the resetting position is relatively small, explaining the power-law
tails in Eq. (\ref{eq:px-tails-alphagt1}). That the MSD is infinite for $1<\alpha<2$ is explained by the fact that
in this range, $\langle \tau^2 \rangle$ is infinite, so that although trajectories on an {\em average} are reset after a
time $\langle \tau \rangle$, there are huge fluctuations around the
average in the {\em actual} time between resets. This feature leads at a given
time $t$ to have a finite probability for the particle
to be at a position $|x| \gg |x_0|$, owing to trajectories that were
last reset in a time of duration substantially longer than $\langle \tau
\rangle$. Such events contribute a
fat-enough tail to $P^{\rm r}_{{\rm
ss}}(x|x_0)$ that the MSD does not have a finite value even at long
times. Invoking a similar argument implies a finite MSD at long times for $\alpha>2$ when $\langle \tau^2\rangle$ is finite.

{\bf First-passage time:}
Let $f^{\rm r}(x_0,T)$ be the first-passage time distribution (FPTD), i.e.,
$f^{\rm r}(x_0,T)dT$ is the probability that the motion starting at
$x=0$ crosses $x_0$ for the first time between times $T$ and $T+dT$.
We have $f^{\rm r}(x_0,T)=-\partial q(x_0,T)/\partial T$, with $q(x_0,T)$ the probability that the motion has not crossed $x_0$ up to
time $T$. The mean first-passage time (MFPT) is $\langle
T\rangle\equiv\int_0^{\infty}dT Tf^{\rm r}(x_0,T)=\widetilde{q}(x_0,0),$
where $\widetilde{q}(x_0,s)$ is the LT of
$q(x_0,T)$, and we have used $q(x_0,\infty)=0$. A renewal theory argument akin to that used for $P^{\rm r}(x,t|x_0,0)$ gives
\be
f^{\rm r}(x_0,T)=\int_0^{T}d\tau
q(x_0,T-\tau)f_{\alpha}(T,T-\tau)f(x_0,\tau),
\label{eq:fxt-gen}
\ee
since a trajectory reaching $x_0$ from $x=0$ for the first time at time $T$ is last reset at an earlier instant $T-\tau; \tau
\in [0,T]$, and has not passed through $x_0$ before that.

\begin{figure}
\includegraphics[width=5cm]{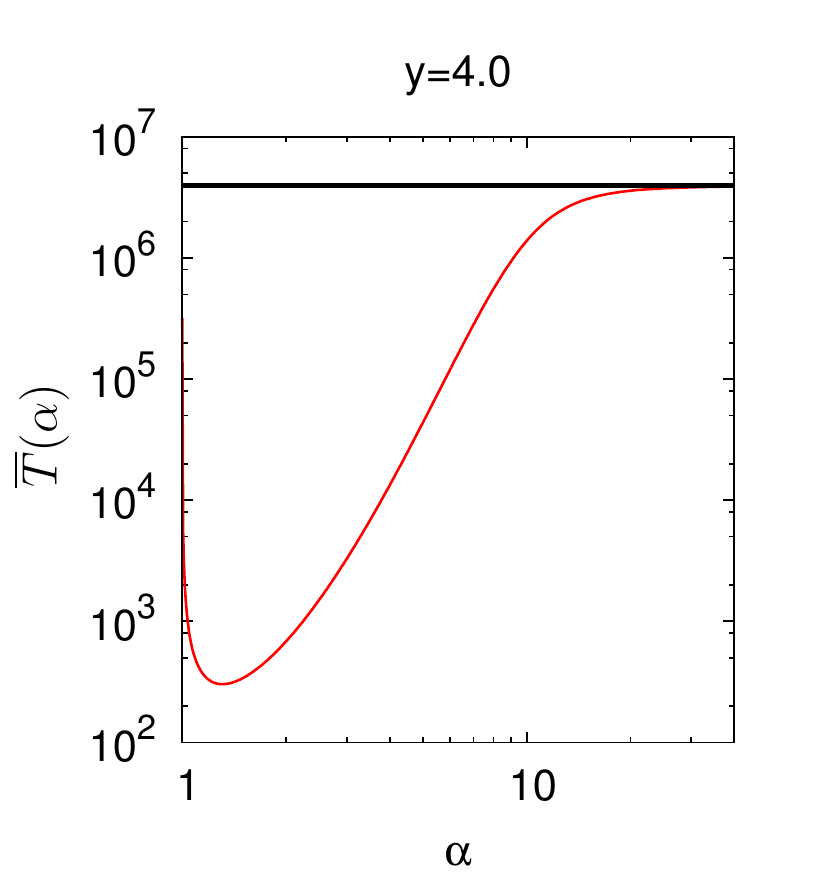}
\caption{$\overline{T}(\alpha)$ versus
$\alpha$, showing the existence of a minimum.}
\label{fig:mfpt}
\end{figure}

Note that in absence of resetting, we have the FPTD $f(x_0,T)=|x_0|/\sqrt{4\pi DT^{3}}\exp[-x_0^{2}/(4DT)]$, thus, $\langle T\rangle=\infty$ \cite{Redner:2007}.
In our case, the existence of a steady state for $\alpha>1$ allows for a finite
MFPT, which we now demonstrate. Let us introduce a dimensionless variable $y\equiv
|x_0|/\sqrt{4D\tau_0}$, given by the ratio of the
distance to the location of desired first passage to the diffusive
length scale in the system. The LT of Eq.
(\ref{eq:fxt-gen}) gives the dimensionless MFPT
$\overline{T}(\alpha) \equiv \langle T\rangle/\tau_0$
as a function of $y \gg 1$ \cite{SM}:
\be
\overline{T}(\alpha)
=\sqrt{\pi}\Big(\frac{\alpha}{\alpha-1}\Big)\Big[ye^{-y^{2}}+\frac{\gamma(\alpha+1/2,y^2)}{y^{2\alpha}}\Big]^{-1}.
\label{eq:Tav-gamma}
\ee
As $\alpha\to\infty$,
$\overline{T}(\alpha\to\infty)=(\sqrt{\pi}/y)\exp(y^{2})$.
The expression for $f_{\alpha>1}$ implies that this limit corresponds to resetting deterministically after every
$\tau_0$ time, so that the FPTD is $re^{-rt}; r\equiv y/(\sqrt{\pi})\exp(-y^{2})$, leading to the form of
$\overline{T}(\alpha\to\infty)$. Figure \ref{fig:mfpt} shows that the MFPT
at a fixed $y$ changes non-monotonically with $\alpha$; The value at which
$\overline{T}(\alpha)$ shows a minimum as a function of
$\alpha$ can be obtained numerically. The existence of a minimum implies
a result relevant both physically and in the context of search
processes in a crowded environment. Namely, for a given distance $|x_0|$ to a fixed target and a given diffusion constant $D$, an optimal $\alpha$ minimizes the time to get to the target for the first time.

Equation (\ref{eq:Tav-gamma}) implies that the MFPT diverges as $\alpha$ approaches unity
from above, and in fact, the MFPT is infinite for $\alpha<1$. This is
because for $\alpha<1$, the long-time behavior is
similar to free diffusion, with the spatial distribution expanding
indefinitely in time. Then, the probability of a typical
trajectory to achieve a first passage through a given location fixed in
space gets smaller with time, and only an atypical one reaches the target, resulting in an infinite MFPT. 

{\bf Conclusions:} We considered the dynamics of a
particle diffusing and resetting to its initial position at random 
times sampled from a power-law $\sim \tau^{-(1+\alpha)}$. Our exact
calculations demonstrated many interesting effects: on tuning $\alpha$
across $1$, the motion at long times crosses over
from being unbounded in time to one that is time
independent even in the absence of boundaries. This behavior may be
contrasted with resetting at exponentially-distributed times that always leads to a time-independent state at long times. A surprising
behavior emerges in the time-dependent spatial distribution around the resetting location for $\alpha<1$: it shows a crossover from a cusp for $\alpha<1/2$ to a divergence for $1/2>\alpha>1$. Although the motion at
long times is time independent for $\alpha>1$, the mean-squared displacement diverges with time
for $1<\alpha<2$, but is time independent for $\alpha>2$. For the mean time to reach for the first time a distant target fixed in space,
we revealed for $\alpha>1$ that there exists of all possible reset strategies
an optimal one corresponding to a particular $\alpha$ that minimizes the mean time.

Our investigations open up many possibilities for future studies. In the
context of search problems, it is interesting to study the time to
reach targets randomly distributed in space by one/many
independent searchers. Such a situation emerges in the context
of animal foraging, where a reset corresponds to returning to the
nest \cite{Giuggioli:2010}. One may further study the effects
of disorder in space due to geographical obstructions/predators that
alter the path of a searcher. To this end, our set-up can be generalized to a motion on a
lattice with every site having as a waiting time a random
variable quenched in space and time. Another interesting follow-up
of our work is to extend it to many-particle interacting systems, and
investigate how dynamics at multiple scales interplays with resetting. Our observed crossovers arise from the non-trivial time dependence of
the probability of last reset, and should be observable in other
systems; Our initial results on interfaces confirm this expectation \cite{Gupta:2016}. 

SG thanks A. C. Barato and L. Giuggioli for discussions, and R. Klages and M. G. Potters for critically reading the manuscript.

\appendix
\begin{widetext}
\section{Derivation of Eqs. (3) and (4) of the main text}
Here, we derive Eqs. (3) and (4) of the main text.
We first note that the Laplace transform (LT) $\widetilde{\rho}(s)\equiv \int_{\tau_0}^\infty d\tau~e^{-s\tau}
\rho(\tau)$ gives for $s\to 0$, $\widetilde{\rho}(s) \approx 1-s\langle\tau\rangle+\langle\tau^{2}\rangle
s^{2}/2$ for $\alpha>2$, $\widetilde{\rho}(s) \approx 1-\langle\tau\rangle
s+as^{\alpha};~a=\Gamma(1-\alpha)\tau_0^{\alpha}$ for $1<\alpha<2$, and
$\widetilde{\rho}(s) \approx 1-|a|s^{\alpha}$ for $\alpha<1$.

As discussed in the main text, $g_n(t);n\ge0$, the probability density for the $n$-th reset at
time $t$, satisfies
$g_n(t)=\int_0^{t}d\tau~\rho(t-\tau)g_{n-1}(\tau);~n\ge1$. Here,
$\int_0^\infty dt g_n(t)=1~\forall~n$, and $g_0(t)=\delta(t)$ accounts for the initial condition $x=x_0$ at $t=0$,
which itself is a reset. An LT operation yields
$\widetilde{g}_n(s)=\widetilde{\rho}(s)\widetilde{g}_{n-1}(s)$, leading
to $\widetilde{g}_n(s)=[\widetilde{\rho}(s)]^{n};~n\ge1$,
on using $\widetilde{g}_0(s)=1$. By definition, we have
$G(t)=\delta(t)+\sum_{n=1}^{\infty}g_n(t)$, whose LT yields
$\widetilde{G}(s)=[1-\widetilde{\rho}(s)]^{-1}$, on using the derived expression
for $\widetilde{g}_n(s)$. \\
\underline{$\alpha>2$:} Here, the final value theorem gives
$G(t\to\infty)=\lim_{s\to0}s\widetilde{G}(s)=1/\langle\tau\rangle=(\alpha-1)/(\alpha \tau_0)$.
The same expression for $G(t)$ also holds for $1<\alpha<2$. \\
\underline{$\alpha<1$}: Here, $\lim_{s\to
0}\widetilde{G}(s)=1/(as^{\alpha})$ yields $G(t\to\infty)\sim
t^{\alpha-1}$. 

Armed with the above results, we now derive Eqs. (3) and (4) of the main text
\subsection{Derivation of Eq. (3) of the main text}
We start with the relation $f_\alpha(t,t-\tau)=\rho_0(\tau)G(t-\tau)$ (see main text),
and the expression $G(t)\sim t^{\alpha-1}$ for large $t$, where $\rho_0(\tau)\equiv \int_\tau^\infty d\tau'\rho(\tau')=(\tau/\tau_0)^{-\alpha};~\tau \ge
\tau_0$. Thus, we have, for $\tau \ge \tau_0$ and for large $t-\tau$ the expression 
\be
f_{\alpha<1}(t,t-\tau)\sim\tau^{-\alpha}(t-\tau)^{\alpha-1}.
\ee
In this case, it is known by a more rigorous treatment that for large $t$, one has \cite{Godreche:2001}
\be
f_{\alpha<1}(t,t-\tau)=\frac{\sin(\pi\alpha)}{\pi}\tau^{-\alpha}(t-\tau)^{\alpha-1},
\ee
which is Eq. (3) of the main text. Note that $\int_0^t d\tau~f_{\alpha<1}(t,t-\tau)=1$.

\subsection{Derivation of Eq. (4) of the main text}
The function $f_{\alpha>1}(t,t-\tau)$ for $t \gg \tau_0$ is given by 
\bea
f_{\alpha>1}(t,t-\tau) & =\begin{cases}
(t/\tau_{0})^{-\alpha}\delta(t-\tau); & \tau=t,\\
(\tau/\tau_{0})^{-\alpha}(\alpha-1)/(\alpha\tau_{0}); & \tau_{0} \le \tau<t-\tau_0,\\
f_{\alpha>1,\tau<\tau_0}(t,t-\tau); & 0<\tau<\tau_{0}.
\end{cases}
\label{eq:fa<1}
\eea
To derive these expressions, note that for $\tau=t,$ the function gives the probability at time $t$ that the last reset was at the initial
instant, which is thus given by $\delta(t-\tau)$ (since we know that the initial condition
itself is a reset) times the probability that no reset appears in an interval of time $t$; the latter probability is given by $\rho_{0}(t)=(t/\tau_{0})^{-\alpha}$. For $\tau_0 \le \tau<t-\tau_0$, the expression for $f_{\alpha>1}(t,t-\tau)$ is derived by using $f_{\alpha>1}(t,t-\tau)=\rho_0(\tau)G(t-\tau)$, and the expression $G(t)=(\alpha-1)/(\alpha \tau_0)$ for large $t$. 

The normalization condition $\int_{0}^{t}d\tau\, f_{\alpha>1}(t,t-\tau)=1$, on using Eq. (\ref{eq:fa<1}), yields
\be
\Big(\frac{t}{\tau_{0}}\Big)^{-\alpha}+\frac{\alpha-1}{\alpha\tau_{0}}\int_{\tau_{0}}^{t-\tau_{0}}d\tau\,\Big(\frac{\tau}{\tau_{0}}\Big)^{-\alpha}+\int_{0}^{\tau_{0}}d\tau\, f_{\alpha>1,\tau < \tau_0}(t,t-\tau)=1.
\ee
The condition $t\gg\tau_{0}$ allows to neglect the first term on the left hand side, so that one has 
\be
\frac{\alpha-1}{\alpha\tau_{0}}\int_{\tau_{0}}^{t}d\tau\,\Big(\frac{\tau}{\tau_{0}}\Big)^{-\alpha}+\int_{0}^{\tau_{0}}d\tau\, f_{\alpha>1,\tau < \tau_0}(t,t-\tau)=1,
\ee
giving 
\be
\int_{0}^{\tau_{0}}d\tau\, f_{\alpha>1,\tau < \tau_0}(t,t-\tau)=1-\frac{\alpha-1}{\alpha\tau_{0}}\int_{\tau_{0}}^{t}d\tau\,\Big(\frac{\tau}{\tau_{0}}\Big)^{-\alpha}.
\ee
To summarize, in the limit $t \gg \tau_0$, one has \cite{Godreche:2001}
\bea
&&f_{\alpha>1,\tau \ge \tau_0}(t,t-\tau)=
\frac{1}{\tau_0}\Big(\frac{\alpha-1}{\alpha}\Big)\Big(\frac{\tau}{\tau_0}\Big)^{-\alpha},
\label{eq:fa>1-limit}
\eea
and  $\int_0^{\tau_0}d\tau
f_{\alpha>1,\tau<\tau_0}(t,t-\tau)=1-\int_{\tau_0}^t
d\tau~f_{\alpha>1,\tau \ge \tau_0}(t,t-\tau)$. Equation (\ref{eq:fa>1-limit}) is Eq. (4) of the main text. 

\section{Derivation of Eq. (5) of the main text}
Here, we derive Eq. (5) of the main text. Equations (2) and (3) of the main text give for $t \gg \tau_0$ the expression
\be
P^{\rm
r}(x,t|x_0,0)=\frac{\sin(\pi\alpha)}{\pi}\int_0^{t}d\tau\,\tau{}^{-\alpha}\Big(t-\tau\Big)^{\alpha-1}\frac{e^{-\frac{(x-x_0)^{2}}{4D\tau}}}{\sqrt{4\pi
D\tau}}.
\ee
Using the transformation $z=t/\tau,$ and the definition of the confluent Hypergeometric
function or the Kummer's function of the second kind $U(a,b,x)$ as
$U(a,b,x)=e^x\int_1^{\infty}dy\, e^{-yx}(y-1)^{a-1}y^{b-a-1}/\Gamma(a)$
\cite{Abramowitz:1972}, we get 
\bea
P^{\rm r}(x,t|x_0,0)&=&\frac{\Gamma(\alpha)\sin(\pi\alpha)e^{-\frac{(x-x_0)^{2}}{4Dt}}}{\pi\sqrt{4\pi
Dt}}U\left(\alpha,\alpha+\frac{1}{2},\frac{(x-x_0)^{2}}{4Dt}\right),
\eea
which is Eq. (5) of the main text.

\section{Derivation of Eq. (7) of the main text}
Here, we provide details on the derivation of Eq. (7) of the main text.
From Eq. (2) of the main text, we get for $t \gg \tau_0$ that
\bea
&&P^{\rm r}(x,t|x_0,0)=\int_{\tau_0}^{t}d\tau\,
f_{\alpha>1,\tau\ge \tau_0}(t,t-\tau)P(x,t|x_0,t-\tau)+ \int_0^{\tau_0}d\tau\, f_{\alpha>1,\tau< \tau_0}(t,t-\tau)P(x,t|x_0,t-\tau)\nonumber \\
&&=\int_{\tau_0}^{t}d\tau\,
f_{\alpha>1,\tau\ge \tau_0}(t,t-\tau)P(x,t|x_0,t-\tau)+ P(x,t|x_0,t-\tau_0)\int_0^{\tau_0}d\tau\, f_{\alpha>1,\tau< \tau_0}(t,t-\tau)\nonumber \\
&&=\int_{\tau_0}^{t}d\tau\,
f_{\alpha>1,\tau\ge \tau_0}(t,t-\tau)P(x,t|x_0,t-\tau)+P(x,t|x_0,t-\tau_0)\Big[1-\int_{\tau_0}^{t}d\tau\,
 f_{\alpha>1,\tau\ge\tau_0}(t,t-\tau)\Big],
\label{eq:px-derivation-alphagt1-ap} 
\eea
where in obtaining the second equality, we have exploited the smallness of $\tau_0$ in approximating the integral in the
second term on the right hand side of the preceding equality.
Using now Eq. (4) of the main text and the
expression for the free
diffusion propagator $P(x,t|x_0,t-\tau)$ in Eq. (\ref{eq:px-derivation-alphagt1-ap}) give
\bea
&&P^{\rm
r}(x,t|x_0,0)=\Big\{1-\frac{1}{\alpha}\Big[1-\Big(\frac{t}{\tau_0}\Big)^{1-\alpha}\Big]\Big\}\frac{e^{-\frac{(x-x_0)^2}{4D\tau_0}}}{\sqrt{4\pi
D \tau_0}}\nonumber \\
&&+\Big(\frac{\alpha-1}{\alpha}\Big)\frac{1}{\sqrt{4\pi
D\tau_0}}\Big[\frac{4D\tau_0}{(x-x_0)^{2}}\Big]^{\alpha-1/2}\Big[\Gamma\Big(\alpha-\frac{1}{2},\frac{(x-x_0)^{2}}{4Dt}\Big)-\Gamma\Big(\alpha-\frac{1}{2},\frac{(x-x_0)^{2}}{4D\tau_0}\Big)\Big],
\label{eq:pxt-agt1-full-eqn-ap}
\eea
where $\Gamma(a,x)$ is the upper incomplete Gamma function. On using the expansion $\Gamma(a,x)=\Gamma(a)\Big[1-x^{a}\exp(-x)\sum_{k=0}^{\infty}x^{k}/\Gamma(a+k+1)\Big];~a\ne0,-1,-2,\ldots,$
Eq. (\ref{eq:pxt-agt1-full-eqn-ap}) gives
\bea
&&P^{\rm
r}(x,t|x_0,0)=\Big\{1-\frac{1}{\alpha}\Big[1-\Big(\frac{t}{\tau_0}\Big)^{1-\alpha}\Big]\Big\}\frac{e^{-\frac{(x-x_0)^2}{4D\tau_0}}}{\sqrt{4\pi
D \tau_0}}\nonumber \\
&&+
\Big(\frac{\alpha-1}{\alpha}\Big)\frac{[4D/(x-x_0)^{2}]^{\alpha-1/2}}{\tau_0^{1-\alpha}\sqrt{4\pi
D}}\gamma\Big(\alpha-\frac{1}{2},\frac{(x-x_0)^{2}}{4D\tau_0}\Big)
 -\Big(\frac{\alpha-1}{\alpha}\Big)\frac{e^{-\frac{(x-x_0)^{2}}{4Dt}}}{\tau_0^{1-\alpha}t^{\alpha-1/2}\sqrt{4\pi
 D}}\sum_{k=0}^{\infty}\frac{\Gamma(\alpha-1/2)\Big[\frac{(x-x_0)^{2}}{4Dt}\Big]^{k}}{\Gamma(\alpha+k+1/2)},
\label{eq:pxt-agt1-full-eqn-1-ap}
\eea
which is Eq. (7) of the main text.
Integrating both sides of the above equation with respect
to $x,$ and using 
$\int_0^\infty dy~y^{a-1}\gamma(b,y)dy=-\Gamma(a+b)/a$ for ${\rm Re}(a)<0, ~{\rm Re}(a+b)>0$,
$\int_0^{\infty}dy\, y^{2k}\exp(-y^{2})=\Gamma(k+1/2)/2$ for $2k>-1,$
and $\sum_{k=0}^{\infty}\Gamma(\alpha-1/2)\Gamma(k+\beta)/\Gamma(\alpha+k+1/2)=\Gamma(\alpha-\beta-1/2)\Gamma(\beta)/\Gamma(\alpha-\beta+1/2)$ \cite{Olver:2010}, it may be checked that $P^{\rm r}(x,t|x_0,0)$ is correctly normalized to 
unity.

The limit $t \to \infty$ of Eq. (\ref{eq:pxt-agt1-full-eqn-1-ap}) yields the steady state:
\be
P^{\rm r}_{{\rm
ss}}(x|x_0)=\Big(\frac{\alpha-1}{\alpha\sqrt{4\pi
D\tau_0}}\Big)\mathcal{G}\Big(\frac{|x-x_0|}{\sqrt{4D\tau_0}}\Big);
\ee
$\mathcal{G}(y)=y^{1-2\alpha}\gamma(\alpha-1/2,y^{2})+e^{-y^2}$.

From Eq. (\ref{eq:pxt-agt1-full-eqn-1-ap}), one obtains the mean-squared
displacement (MSD) as a function of time as 
\bea
&&\langle(x-x_0)^2\rangle(t)
=2D\tau_0\Big(\frac{\alpha-1}{\alpha(\alpha-2)}\Big)\Big[1-\Big(\frac{t}{\tau_0}\Big)^{2-\alpha}\Big]+2D\tau_0\Big\{1-\frac{1}{\alpha}\Big[1-\Big(\frac{t}{\tau_0}\Big)^{1-\alpha}\Big]\Big\}.
\label{eq:MSD-exact-ap}
\eea
For $\alpha>2$, the MSD
converges to a finite constant:
\bea
\langle (x-x_0)^2 \rangle (t \to \infty)
&=&2D\tau_0\Big(\frac{(\alpha-1)^2}{\alpha(\alpha-2)}\Big);~\alpha>2.
\label{eq:agt2msd-ap}
\eea
On the other hand, for $\alpha$ in the range $1<\alpha<2$, the MSD diverges with time
as 
\be
\langle (x-x_0)^2 \rangle (t)\approx C_1
t^{2-\alpha}~{\rm for~large~}t;~1<\alpha<2,
\label{eq:alt2msd-ap}
\ee
with the finite constant $C_1$ obtained from Eq. (\ref{eq:MSD-exact-ap}).

In Fig. \ref{fig:px-comparison}, we show a comparison of
numerical simulation data for the spatial distribution at long times with
our analytical result, namely, Eq. (\ref{eq:pxt-alphalt1-exact-1}) for the time-dependent
distribution for $\alpha<1$, and Eq. (\ref{eq:ss-distr}) for the steady state distribution for
$\alpha>1$, for representative values of $\alpha$, thereby
demonstrating an excellent agreement.

\begin{figure}
\centering
\includegraphics[width=10cm]{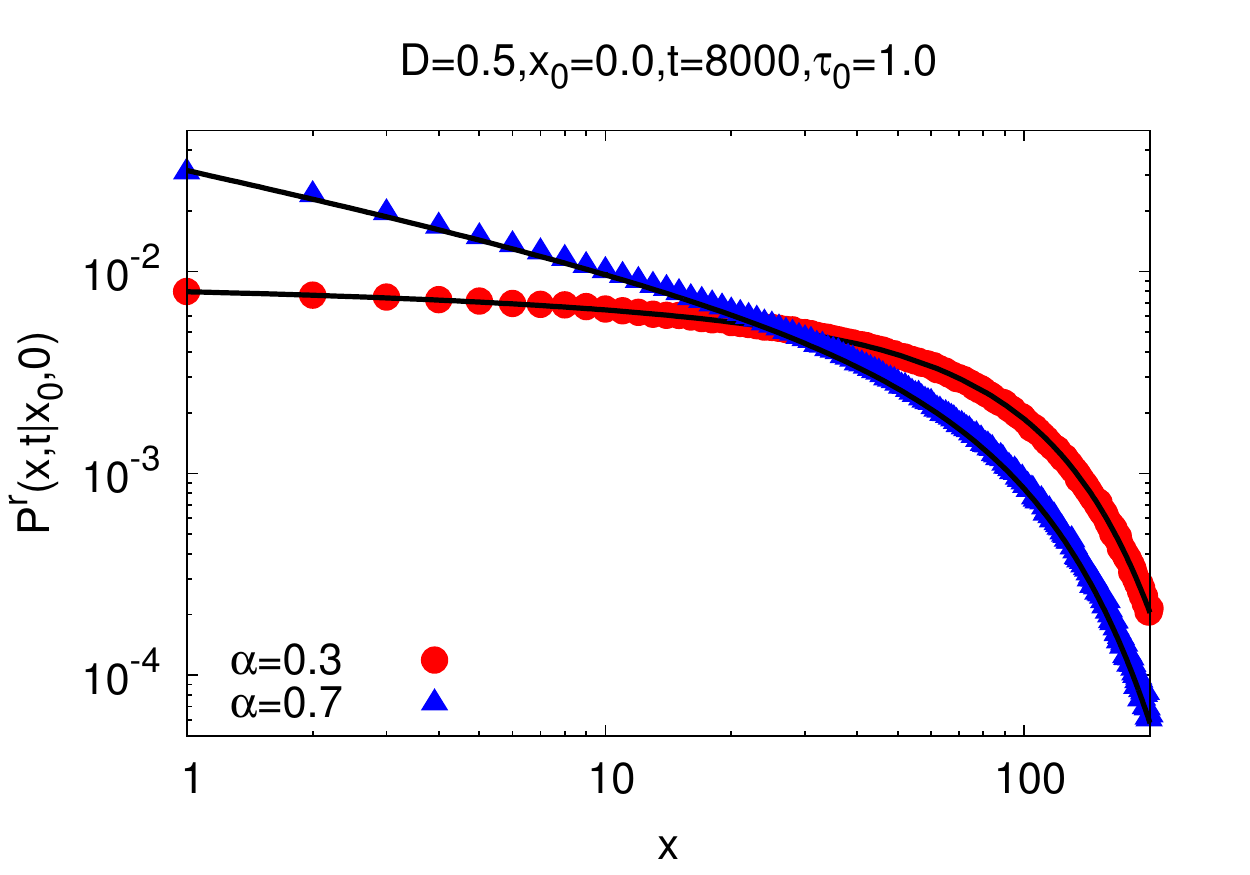}
\includegraphics[width=10cm]{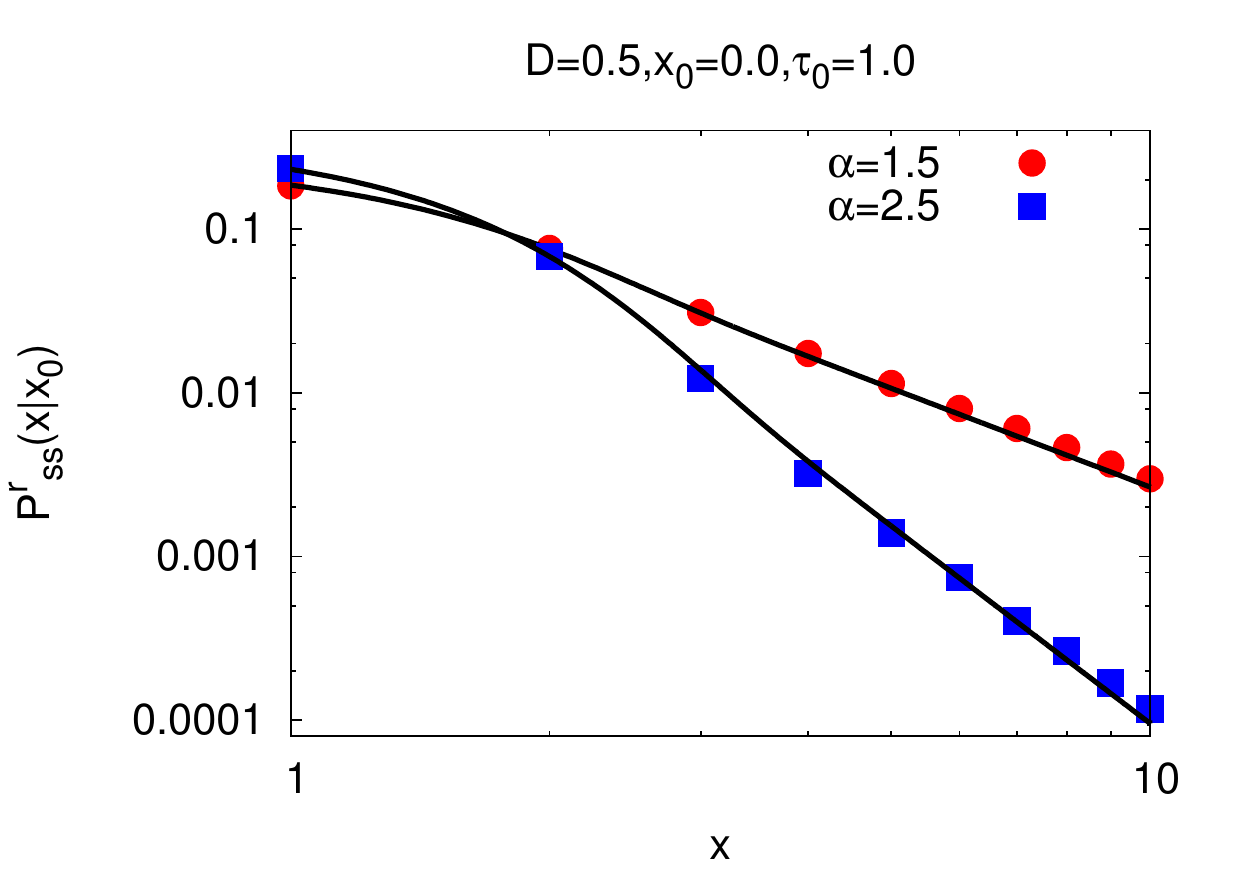}
\caption{(Color online) For representative values of $\alpha$ smaller
and larger than unity, the figures show a comparison of numerical
simulation data (points) for the spatial distribution with analytical
results (lines), namely, Eq. (\ref{eq:pxt-alphalt1-exact-1}) for the time-dependent
distribution for $\alpha<1$, and Eq. (\ref{eq:ss-distr}) for the steady state distribution for
$\alpha>1$.}
\label{fig:px-comparison}
\end{figure}

\section{Derivation of Eq. (11) of the main text}
Here, we give details on the derivation of Eq. (11) of the main text.
From Eq. (10) of the main text, we get for $T \gg \tau_0$ that
\bea
&&f^{\rm r}(x_0,T)=\int_{\tau_0}^{T}d\tau\,
q(x_0,T-\tau)f_{\alpha>1,\tau \ge \tau_0}(T,T-\tau)f(x_0,\tau)  + \int_0^{\tau_0}d\tau\,q(x_0,T-\tau)f_{\alpha>1,\tau < \tau_0}(T,T-\tau)f(x_0,\tau)\nonumber \\
&&=\int_{\tau_0}^{T}d\tau\,
q(x_0,T-\tau)f_{\alpha>1,\tau \ge \tau_0}(T,T-\tau)f(x_0,\tau)+q(x_0,T-\tau_0)f(x_0,\tau_0) \int_0^{\tau_0}d\tau\,f_{\alpha>1,\tau < \tau_0}(T,T-\tau), \nonumber \\
&&=\int_{\tau_0}^{T}d\tau\,
q(x_0,T-\tau)f_{\alpha>1,\tau \ge \tau_0}(T,T-\tau)f(x_0,\tau)+q(x_0,T-\tau_0)f(x_0,\tau_0)\Big(1-\int_{\tau_0}^{T}d\tau\,
 f_{\alpha>1,\tau \ge \tau_0}(T,T-\tau)\Big), 
\label{eq:qeq-1-agt1-1-ap}
\eea 
where in obtaining the second equality, we have assumed the smallness of
$\tau_0$ in approximating the integral in the second term on the right
hand side of the preceding equality. We considered $T \gg \tau_0$ in the above derivation, which is the limit in
which we have an expression for $f_{\alpha>1}(T,T-\tau)$, a crucial
ingredient in the computation of the MFPT, and which in turn implies that
the distance $|x_0|$ to the target through which the first-passage is
desired satisfies $|x_0| \gg \sqrt{2D\tau_0}$. We finally have 
\bea
&&f^{\rm r}(x_0,T)=\Big(\frac{\alpha-1}{\alpha}\Big)\frac{1}{\tau_0}\frac{|x_0|}{\sqrt{4\pi D}}\int_0^{T}d\tau\, q(x_0,T-\tau)\frac{e^{-x_0^{2}/(4D\tau)}}{\tau^{3/2}(\tau/\tau_0)^{\alpha}}+(B-A)q(x_0,T-\tau_0), 
\label{eq:qeq-1-agt1-ap}
\eea
where
\bea
A
\equiv\Big(\frac{\alpha-1}{\alpha}\Big)\frac{|x_0|}{\tau_0\sqrt{4\pi D}}\int_0^{\tau_0}d\tau\,\frac{e^{-x_0^{2}/(4D\tau)}}{\tau^{3/2}(\tau/\tau_0)^{\alpha}}
=\Big(\frac{\alpha-1}{\alpha}\Big)\frac{1}{\tau_0\sqrt{\pi}}\Big(\frac{4D\tau_0}{x_0^{2}}\Big)^{\alpha}\Gamma\Big(\alpha+\frac{1}{2},\frac{x_0^{2}}{4D\tau_0}\Big),
\eea
and
\bea
B&\equiv&\Big[1-\Big(\frac{\alpha-1}{\alpha}\Big)\int_{\tau_0}^{T}d\tau\,\frac{1}{(\tau/\tau_0)^{\alpha}}\Big]\frac{|x_0|}{\sqrt{4\pi
D\tau_0{}^{3}}}e^{-x_0^{2}/(4D\tau_0)}\nonumber \\
&=&\Big(1-\Big(\frac{\alpha-1}{\alpha}\Big)\frac{1-(T/\tau_{0})^{1-\alpha}}{\alpha-1}\Big)\frac{|x_{0}|}{\sqrt{4\pi D\tau_{0}{}^{3}}}\exp[-x_{0}^{2}/(4D\tau_{0})]\nonumber \\
&\approx&\Big(\frac{\alpha-1}{\alpha}\Big)\frac{|x_0|}{\sqrt{4\pi
D\tau_0^{3}}}e^{-x_0^{2}/(4D\tau_0)},
\eea
where $T \gg \tau_0$ and $\alpha >1$ allow to neglect the term
$(T/\tau_{0})^{1-\alpha}$ in the second equality.

The Laplace transform (LT) of Eq. (\ref{eq:qeq-1-agt1-ap}), and $q(x_0,0)=1$ yield
\bea
&&\widetilde{q}(x_0,s) 
=\Big[s+(B-A)e^{-s\tau_0}+\Big((\alpha-1)/\alpha\Big)\tau_0^{\alpha-1}2^{\alpha+1/2}\Big(D^{\alpha/2-1/4}/(\sqrt{\pi}|x_0|^{\alpha-1/2})\Big)
s^{\alpha/2+1/4}K_{\alpha+1/2}(|x_0|\sqrt{s}/\sqrt{D})\Big]^{-1},
\nonumber \\
\eea
on using that the LT of
$\exp[-x_0^{2}/(4D\tau)]/[\tau^{3/2}(\tau/\tau_0)^{\alpha}]$ is
$\tau_0^{\alpha}2^{\alpha+\frac{3}{2}}(Ds/x_0^{2})^{\alpha/2+1/4}K_{\alpha+\frac{1}{2}}\left(|x_0|\sqrt{s}/\sqrt{D}\right)$,
with $K_{\nu}(x)$ being the modified Bessel function of the second kind.
Here, we have also used that $q(x_0,T-\tau)=0$ for $\tau>T$, and the result
that the LT of $f(t-a)u(t-a)$, with $u(t)$ being the Heaviside step
function, equals $e^{-as} F(s)$, where $F(s)$ is the LT of $f(t)$. Now,
the MFPT is given by $\langle T\rangle=\widetilde{q}(x_0,0)$. 
Then, for small $s,$ using $K_{\nu}(x)=(\Gamma(\nu)/2)(2/x)^{\nu},$ we
have in terms of $y\equiv |x_0|/\sqrt{4D\tau_0} \gg 1$ the dimensionless MFPT $\overline{T}(\alpha)
\equiv \langle T\rangle/\tau_0$ given by
\be
\overline{T}(\alpha)
=\sqrt{\pi}\Big(\frac{\alpha}{\alpha-1}\Big)\Big[ye^{-y^{2}}+\frac{\gamma\Big(\alpha+\frac{1}{2},y^2\Big)}{y^{2\alpha}}\Big]^{-1},
\ee
which is Eq. (11) of the main text. 

In Fig. \ref{fig:fpt-comparison}, we show a comparison of numerical
simulation data for the MFPT with the analytical
result given by Eq. (\ref{eq:Tav-gamma}). A good agreement is
evident from the figure for $x_0 \gg \sqrt{2D\tau_0}$. 

\begin{figure}
\centering
\includegraphics[width=10cm]{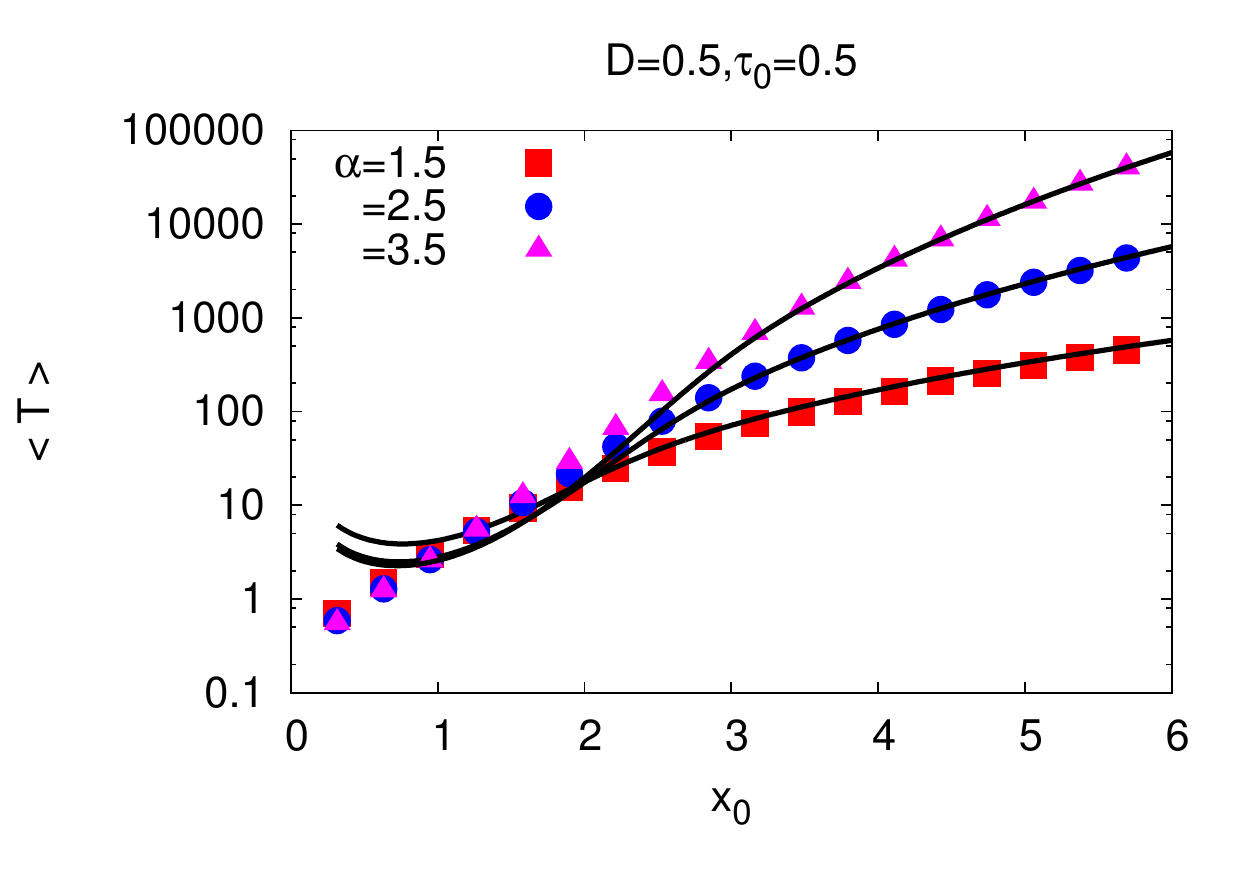}
\caption{(Color online) For representative values of $\alpha$, the
figure shows a comparison of numerical
simulation data (points) for the MFPT with analytical
results (lines) given by Eq. (\ref{eq:Tav-gamma}).}
\label{fig:fpt-comparison}
\end{figure}

\end{widetext}


\begin{thebibliography}{99}
\bibitem{norm-note}Integrating Eq. (\ref{eq:pxt-gen}) over $x$, using $\int_0^t d\tau
f_\alpha(t,t-\tau)=1$ and normalization of $P(x,t|x_0,0)$, it is checked that $P^{{\rm r}}(x,t|x_0,0)$ is
normalized to unity.
\end{thebibliography}
\end{document}